\documentstyle[12pt,epsfig]{article}                                
\begin{document}

\titlepage                  

\begin{center}
\begin{Large}
\begin{bf}

  Derivation and correlation between PYTHAGORAS (569 -479 B.C) and MATHIEU'S (1868) equation: Spectral nature between MATHIEU's and MODIFIED MATHIEU's equation

\vspace{1.0cm}

\end{bf}
\end{Large}

 Biswanath Rath 

\end{center}

\vspace{0.1cm}

\begin{it}
 Department of Physics,
 Maharaja Sriram Chandra Bhanj Deo University,
 Takatpur, Baripada -757003, Odisha, India.
e.mail:biswanathrath10@gmail.com

\vspace{0.1cm}

\end{it}

$\bf{Abstract:}$
We derive Pythagoras theorem. From the Pythagoras theorem, we also derive 
Mathieu's equation via modified Mathieu's equation.
A spectral comparison has been carried out between modified mathieu's equation and Mathieu's equation. Apart from this, we also present discrete bound states 
corresponding to modified Mathieu's equation of a quantum rectangular type of model potential. 
\vspace{1.0cm}

\begin{bf}

\hspace{0.10cm}


\hspace{0.10cm}

01A20: History of mathematics in ancient Greece and Rome.

\hspace{0.10cm}

15A18: Eigenvalues, singular values, eigenvector. 

\hspace{0.10cm}

\end{bf}

\vspace{0.1cm}

\hspace{0.5cm} \noindent\rule{3.0in}{0.4pt}

Correspondence: biswanathrath10@gmail.com 

\begin{bf}
1.Introduction
\end{bf}

Nearly 4000 years ago, a famous Greek mathematician cum philosopher named " Pythogoras" proposed that if sum of  squares of two sides of a triangle becomes squares of the third side, then the triangle must be "right angle triangle".
 Mathematically 
\begin{equation}
a^{2}+b^{2}=c^{2}
\end{equation}
where $a,b$ and $c$ are the sides of the triangle.

\vspace{1.0cm}
\begin{table}
\begin{center}
\begin{tabular}{c  } \\
\includegraphics[width= 0.5\textwidth]{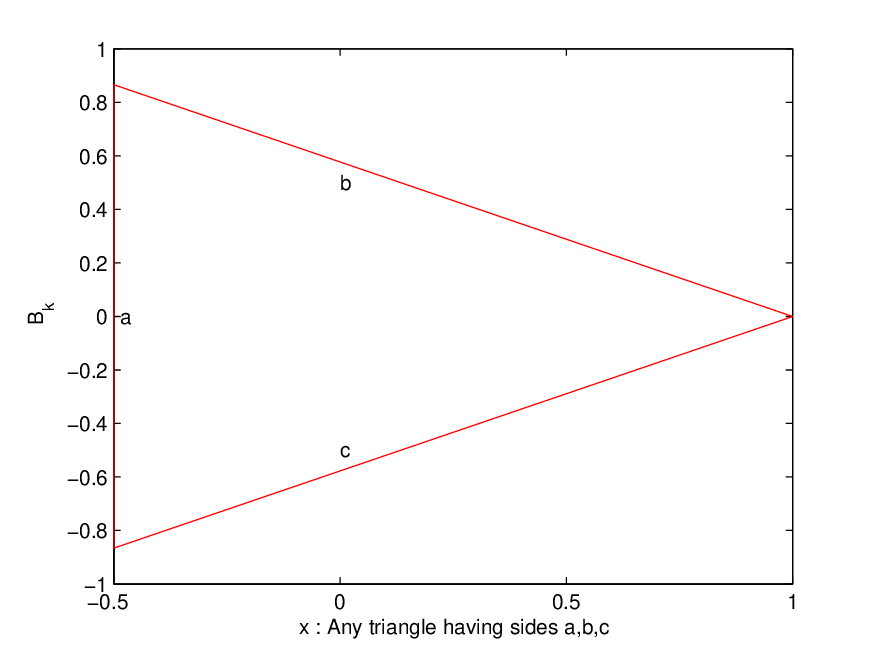} \\
\hspace{0.1cm}Fig.1. Any triangle    \\
\end{tabular}
\end{center}
\end{table}

Further while studying vibrations of string nearly one and a half century ago a 
French mathematician named Mathieu proposed that vibrations can best be described by a differential equation of the form

\begin{equation}
 - U_{xx} + 2q \cos(2x) U = \lambda U
\end{equation}

where $'q$ is a real number and $\lambda $ a constant.
If one compares the time interval between the two, there is a very large gap. 
On one hand one is pure mathematics dealing with geometry while the other is
 physical examples related to mathematics of music. In fact, we find both of two ae intimately related. Since Mathieu's equation is a later one, we prefer to 
derive the same from the former one. Prior to this, we first prove the 'Pythagoras theorem '.

\begin{bf}
2a.A difficult proof of  Pythagoras theorem.
\end{bf}

Let us define the following 
\begin{equation}
\frac{a}{c}= \frac{1}{\cosh(sin(x))}
\end{equation}
\begin{equation}
\frac{b}{c}= \tanh(sin(x))
\end{equation}
Hence it is to show that 
\begin{equation}
b=a \sinh(sin(x))
\end{equation}
Hence using the relation 

\begin{equation}
(\cosh(sin(x)))^{2} = 1 + (\sinh(sin(x)))^{2} 
\end{equation}

\begin{equation}
= 1+\frac{b^{2}}{a^{2}}=\frac{a^{2}+b^{2}}{a^{2}}
\end{equation}
\begin{equation}
=[a^{2}+ b^{2}]/[\frac{c}{\cosh(x)})^{2}
\end{equation}
This implies 
\begin{equation}
1=\frac{a^{2}+b^{2}}{c^{2}} \Longrightarrow a^{2}+b^{2}=c^{2}
\end{equation}

Hence the theorem is proved.

\begin{bf}
2b.A simpler  proof of  Pythagoras theorem.
\end{bf}

Here we replace $\sin(x)\rightarrow x$.
Hence the above equations are written as 
\begin{equation}
\frac{a}{c}= \frac{1}{\cosh(x)}
\end{equation}

and 

\begin{equation}
\frac{b}{c}= \tanh(x)
\end{equation}
Hence it is to show that 
\begin{equation}
b=a \sinh(x)
\end{equation}

Hence using the relation 

\begin{equation}
(\cosh(x))^{2} = 1 + (\sinh(x))^{2} 
\end{equation}

\begin{equation}
= 1+\frac{b^{2}}{a^{2}}=\frac{a^{2}+b^{2}}{a^{2}}
\end{equation}

Now substituting $a$ in terms of $ sech(x)$(only in denominator) we have

\begin{equation}
(\cosh(x))^{2} = (a^{2}+b^{2})/a^{2}
\end{equation}

\begin{equation}
=[a^{2}+ b^{2}]/[\frac{c}{\cosh(x)})^{2}
\end{equation}
This implies 
\begin{equation}
1=\frac{a^{2}+b^{2}}{c^{2}} \Longrightarrow a^{2}+b^{2}=c^{2}
\end{equation}

Hence the theorem is proved.

\vspace{1.0cm}
\begin{table}
\begin{center}
\begin{tabular}{c  } \\
\includegraphics[width= 0.5\textwidth]{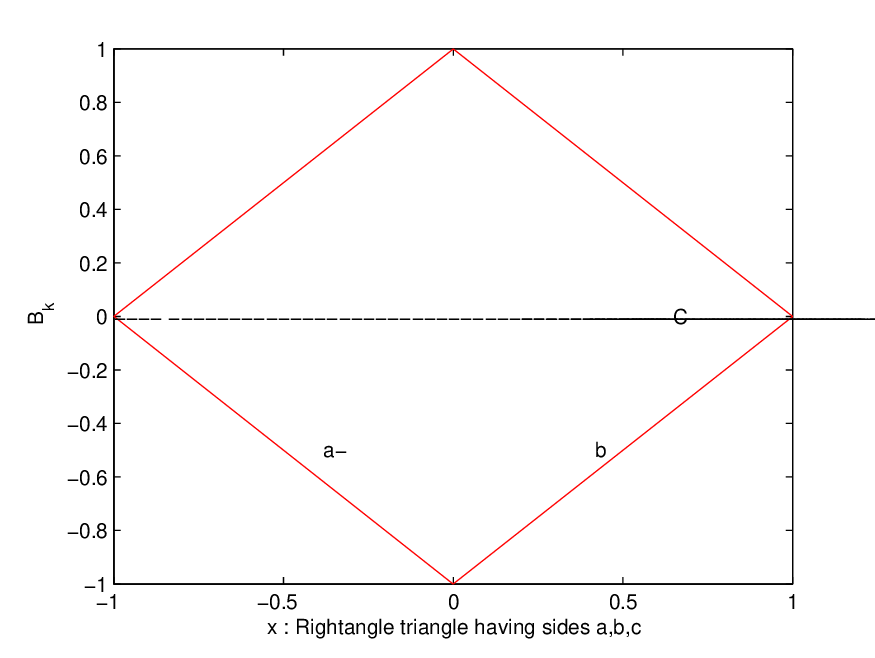} \\
\hspace{0.1cm}Fig.2. Rightangle triangle   \\
\end{tabular}
\end{center}
\end{table}

\begin{bf}
3.Derivation of  Mathieu's equation
\end{bf}

 Let us consider the relation 
\begin{equation}
[\frac{a^{2}}{c^{2}} + (\frac{b}{c})^{2}=1=(\frac{1}{\cosh(x)})^2+(\tanh(x))^{2}
\end{equation}

Hence we have the relation

\begin{equation}
(\cosh(x))^{2}-(\sinh(x))^{2}=1
\end{equation}

Letus define a new potential as 

\begin{equation}
V(x)=[\frac{\cosh(x)}{L}]^{2}-(L*\sinh(x))^{2}
\end{equation}

where $L$ is greater than unity. The above potential is written as 

\begin{equation}
V(x)= 2A \cosh(2x) + B
\end{equation}
where $A=[\frac{1}{L^{2}}-L^{2}]/4$ and $B=[\frac{1}{L^{2}}+L^{2}]/2$.
Hence the corresponding Hamiltonian can be written as 
\begin{equation}
H=p^{2} + V(x)
\end{equation}
In its differential form, this is written as 

\begin{equation}
U_{xx} + (\lambda - 2A \cosh(2x)]U = 0
\end{equation}
this is known as 
\begin{bf}
modified Mathieu's equation.
\end{bf}
where $\lambda=E- B$ and $U_{xx}=  \frac{d^{2} U}{d x^{2}}$.
 the potential if one substitute $x=ix$ , then above equation becomes Mathieu's 
equation
\begin{equation}
U_{xx} + (\lambda - 2A \cos(2x)) U =0
\end{equation}
which is known as 

\begin{bf}
 Mathieu's equation.
\end{bf}

Hence we find Mathieu's equation and Pythagoras rlation are intimately  related
to each other. In other words Mathieu's equation can be derived from Pythagoras  theorem.
\

\begin{bf}

4. Comparison of spectra between Mathieu's equation and modified  Mathieu's equation.
\end{bf}

Here we consider the scale of Abramowitz and Stegun[3] and compare the spectra between the two in fig-2.

\vspace{1.0cm}
\begin{table}
\begin{center}
\begin{tabular}{c c } \\
\includegraphics[width= 0.5\textwidth]{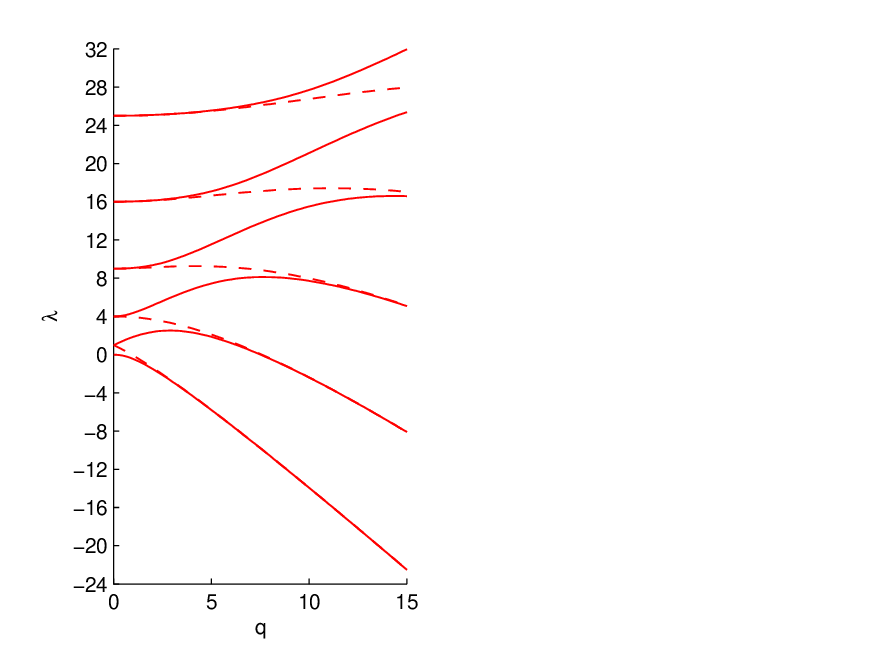} &  
\includegraphics[width= 0.5\textwidth]{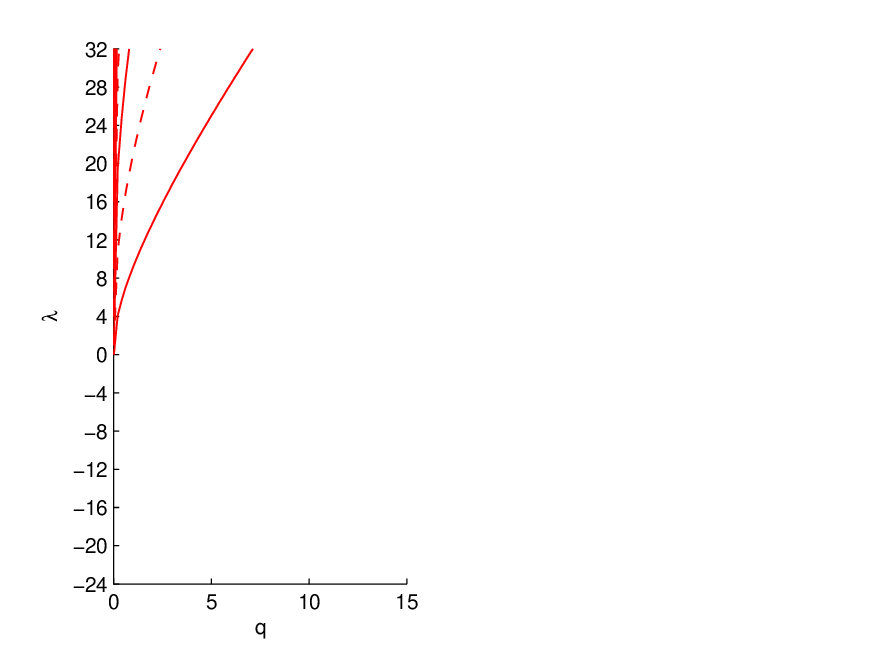}   \\
\hspace{0.1cm}Fig.2.LHS " Mathieu" ; RHS"Modified Mathieu"   \\
\end{tabular}
\end{center}
\end{table}

\begin{bf}

5. Modified Mathieu's equation : Potential well
\end{bf}

Here, we apply Modified Mathieu's equation to study spectral nature of a 
rectangular type of well as in fig 3.

\vspace{1.0cm}
\begin{table}
\begin{center}
\begin{tabular}{c  } \\
\includegraphics[width= 0.5\textwidth]{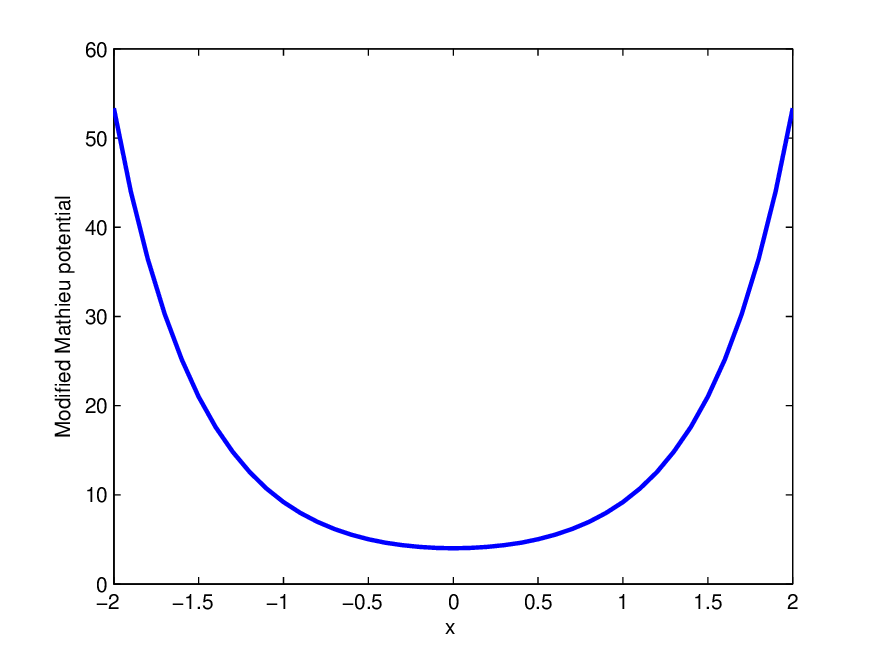} \\
\hspace{0.1cm}Fig.2. Potential nature    \\
\end{tabular}
\end{center}
\end{table}

The corresponding phase-portrait is also closed  in nature(see fig-3)[4]

\vspace{1.0cm}
\begin{table}
\begin{center}
\begin{tabular}{c  } \\
\includegraphics[width= 0.5\textwidth]{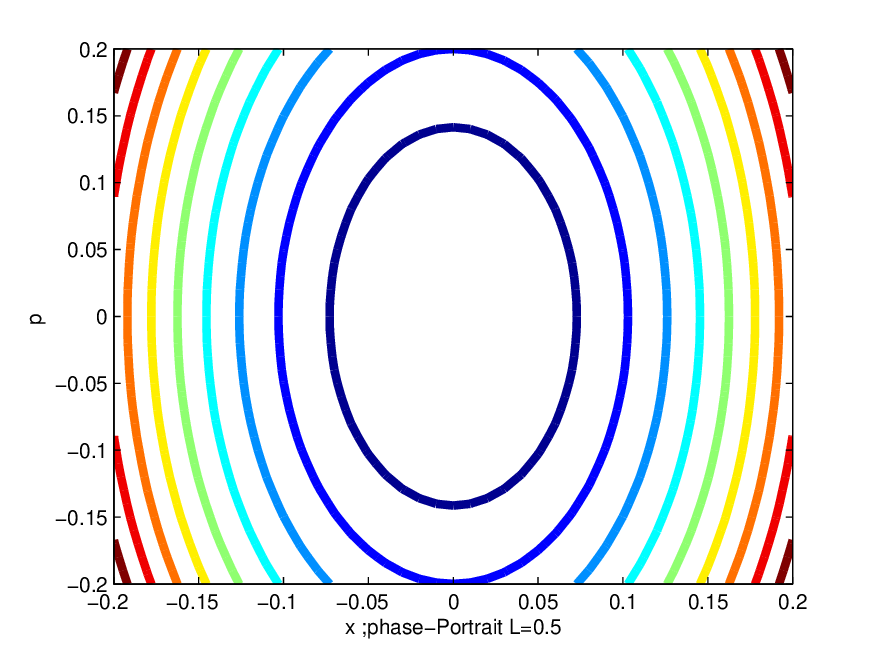} \\
\hspace{0.1cm}Fig.3.Phase-portrait natuure    \\
\end{tabular}
\end{center}
\end{table}

Quantum mechanically corresponding wave functions is written as[2] 
\begin{equation}
\Psi(x)= c_{1} \psi_{1}(ix) + c_{2} \psi_{2}(ix)
\end{equation}
where $\psi(ix) ; \psi_{2}(ix)$ can be obtained from Mathieu's solution with $x\rightarrow ix$[2].

Using suitable Matlab code , we present the first four wave function as given in  fig-3.

\vspace{1.0cm}
\begin{table}
\begin{center}
\begin{tabular}{c  } \\
\includegraphics[width= 0.5\textwidth]{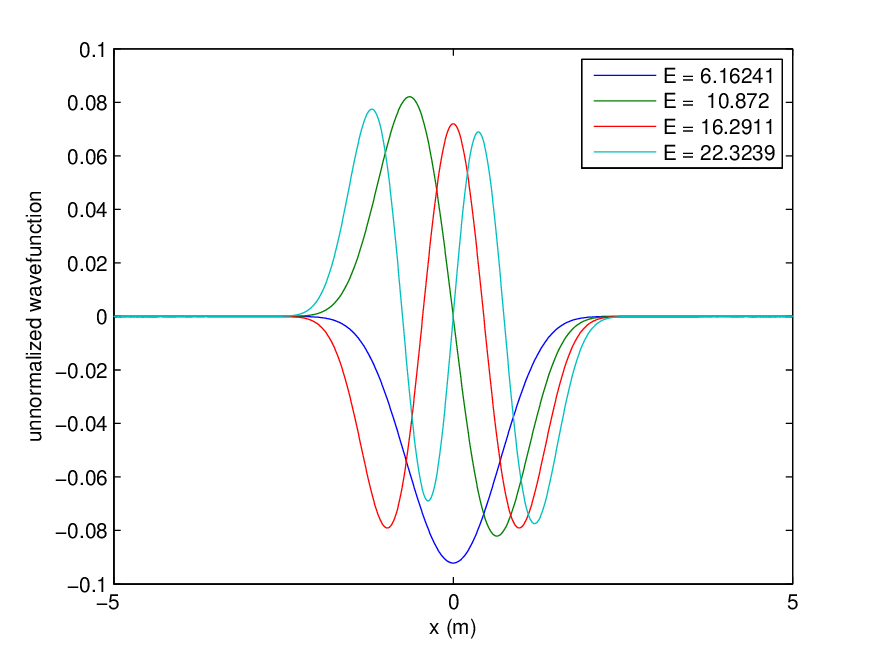} \\
\hspace{0.1cm}Fig.4.wave function nature    \\
\end{tabular}
\end{center}
\end{table}

In table-1, we present the first four energy levels.

\begin{table}[htbp]
\begin{center}

\vspace{1.0cm}

TABLE.I: First four energy levels 

\begin{tabular}{ c c } \\ \hline

n & Energy \\  \hline \hline  
0 & 6.162 41     \\
1 & 10.872    \\
2 & 16.291 1  \\
3 & 22.323 9  \\   \hline \hline  
\end{tabular} 
\end{center}
\end{table}

\begin{bf}
5.Conclusion
\end{bf}

We formulate a modified Mathieu's equation and study its classical and non-classical  nature. Model formlation is suitable for bound states calculation for $L\ll 1).$ 

\pagebreak

\begin{bf}
Author's contribution:
\end{bf}

B.Rath: formulation,computation,writing,finalization.

\begin{bf}
Conflict of interest
\end{bf}

Author declares there is no conflict of interest.

\begin{bf}
DATA AVAILABILITY
\end{bf}

No additional data is required . All the datas  included in this paper are 
sufficient.

\begin{bf}
Declaration
\end{bf}

Present paper is a modified version of arxiv paper.

\end{document}